\documentclass[fleqn]{article}
\usepackage[dvips]{graphicx}
\author{S.Tsereteli, V.Bochorishvili, M.Samkharadze, G. Khoriauli}
\date{}
\title{\large \bf Study of cosmogenic radiocarbon concentration
variations in the Earth's atmosphere during solar activity of
Maunder minimum (1645-1715) }
\begin{document}
\maketitle
\begin{center}
 Tbilisi State University, Georgia, 0128. Tbilisi, Chavchavadze av. 3
\\ Email: bochorishvili@ictsu.tsu.edu.ge
\begin{abstract}
 \hspace {\parindent} Variations of radiocarbon concentration have
  been studied in annual rings for the last 350 years $(1600-1950)$
  on the basis of our experimental research using methods of spectral analysis. From
this interval of time special attention is paid to the so-called
period of the Maunder minimum $(1645-1715)$ of solar activity. In
the experimental series of corresponding period  two types of
periodicity are revealed: $\approx20$ year and $\approx8$ year.
\end{abstract}
\end{center}
\section { Introduction}
 \hspace {\parindent}
The investigation of the dynamics of solar activity and regularity
of its cyclicing is of great importance for   the  astrophysics.
Correspondingly the study of those intervals when solar activity
is sharply changed is of particular significance.  Examples  of
such interesting periods are: 1100-1250 yy "maximum of Middle
Ages", in 1282-1342 yy "Volf minimum", 1416-1534 yy "Shperer
minimum", 1645-1715 yy "Maunder minimum".

The period of the Maunder minimum  has attracted an additional
attention because it puts forward a lot of new and unexpected
problems. Indeed,  for a long time scientists consider spots on
the Sun surface as main characteristics of the solar activity.
Therefore, the notes about the  disappearance of the sunspot is
reliable, but due to the absence of regular observations it is
difficult to determine level of solar activity in the mentioned
period.

It should be noted that the anomalies  are found in differential
rotation of the Sun surface at the beginning of the Maunder
minimum.  According of the different authors, it might indicate
that global characteristics of the inner parts of the Sun change
to such at extant that even the "dynamo" mechanisms is stopped. In
this connection the following questions arise: how to characterize
the Sun activity and the radioactive state in the Earth's space in
that epoch? What is main processes which cause such a drastic fall
of solar activity? To what extent our insight into the solar
activity correct in general?

For solving mentioned problems physicist direct their special
attention to cosmogenic isotopes $^{14}C$ $(T_{1/2}=5730\ year)$,
$^{10}Be$ $(T_{1/2} =1,5 \cdot 10 ^{6}\ year)$. The study of these
isotopes may give a lot of information about the dynamics of solar
activity.

At present the most complete data about the solar activity and the
intensity of cosmic rays are obtained by means of the above
mentioned cosmogenic isotopes ($ ^{14}C$ and $^{10}Be$) formed in
the Earth's atmosphere. The main peculiarity of these isotopes and
advantage as compared with others is the fact that there exist
objects in nature  (e. g. tree rings, Arctic and Antarctic ices,
algae with long life time, etc.) which can "fix" i.e. "memorize"
annual concentration of the mentioned isotopes, during period that
prevails it's half-life time.

\section{Radiocarbon Concentration during the Maunder Minimum }
\hspace {\parindent}To study the influence of solar activity
(modu1ation) in the Earth atmosphere on radiocarbon concentration
during the Maunder minimum we use experimental data obtained at
the laboratory of nuclear physics of the TSU \cite{stradiocarboni,
seminari83, seminari88} (Fig. 1). First of all, it should be noted
that experimental data cover the Maunder minimum itself (1645-1720
yy) (Fig. 2) and its neighboring periods (1600-1645 and 1715-1800
yy) too. The measurements in \cite{stradiocarboni} were performed
with one-year step, i.e. for every calendar year $^{14}C$
concentration is measured in the corresponding tree ring. The
experimental samples of tree rings were obtained in Lithuania.

The value of $^{14}C$ concentration in relation to the
International standard scale is calculated by the following
formula
$$\Delta^{14}C_{0}=\frac{N_{s}-N_{st}}{N_{st}-N_{b}} ,$$
where  $N_{s}$, $N_{st}$,$N_{b}$ represents  measured activities
of the sample (s), standard (st) and background (b), respectively.
Experimental error does not exceed 0.2-0.3\%. In order to extract
variations of $^{14}C$ concentration originated only from the
"cosmic" sources, it is necessary to exclude the influence of the
Earth's magnetic field.  This influence has been widely discussed
and to date the most accurate estimation is that: global change of
geomagnetic field with period of 10 thousand years causes about
10\% change in $^{14}C$ data. The following expression accounts
for the geomagnetic field effect in $^{14}C$ concentration
$$\Delta^{14}C(t)=\Delta^{14}C_{0}(t)-4,27-4,97\cdot\sin[\frac{2\pi}{t+7388}] ,$$
where t is   current calendar year  (t is negative B.C. ).

As we mentioned above  the effective tool  for studying the solar
activity dynamics during the Maunder minimum is to reveal the
cyclings in $^{14}C$ time dependence (i.e. periodicities in
experimental data given in Fig.2). For this purpose the method of
spectral analysis has been used
 \cite{Tsdjenkinsi}. The calculated power spectrum is shown in
Fig.3. According to Fig.3 the main peak at frequency 0.05 lies in
high reliability interval (95\%)  (corresponding period
$\approx20$ year). Small peak is observed at frequency 0.12
(corresponding period $\approx8$ year) though it's reliability  is
comparatively low.
\section {conclusions:}
\begin{enumerate}
\item  Using the method of spectral analysis we have revealed
variations in cosmogenic radiocarbon concentration in the Earth
atmosphere during the Maunder minimum. \item Periodicities of two
types have been revealed in experimental data. \item The reason of
the  mentioned periodicities is the solar activity. This means
that in the Maunder minimum period modulating action of the Sun
was weak, nevertheless cycling is preserved. It can be also said
that in period of the Maunder minimum the mechanism of solar
dynamo still works. Reduction of the number of the sunspots may be
caused by the coincidence of the minima of any other long-term
periodicities at the moment of the Maunder minimum.
\end{enumerate}

\begin{figure}[h!bpt]
\begin {center}
\includegraphics[angle=270, width=\textwidth  ]{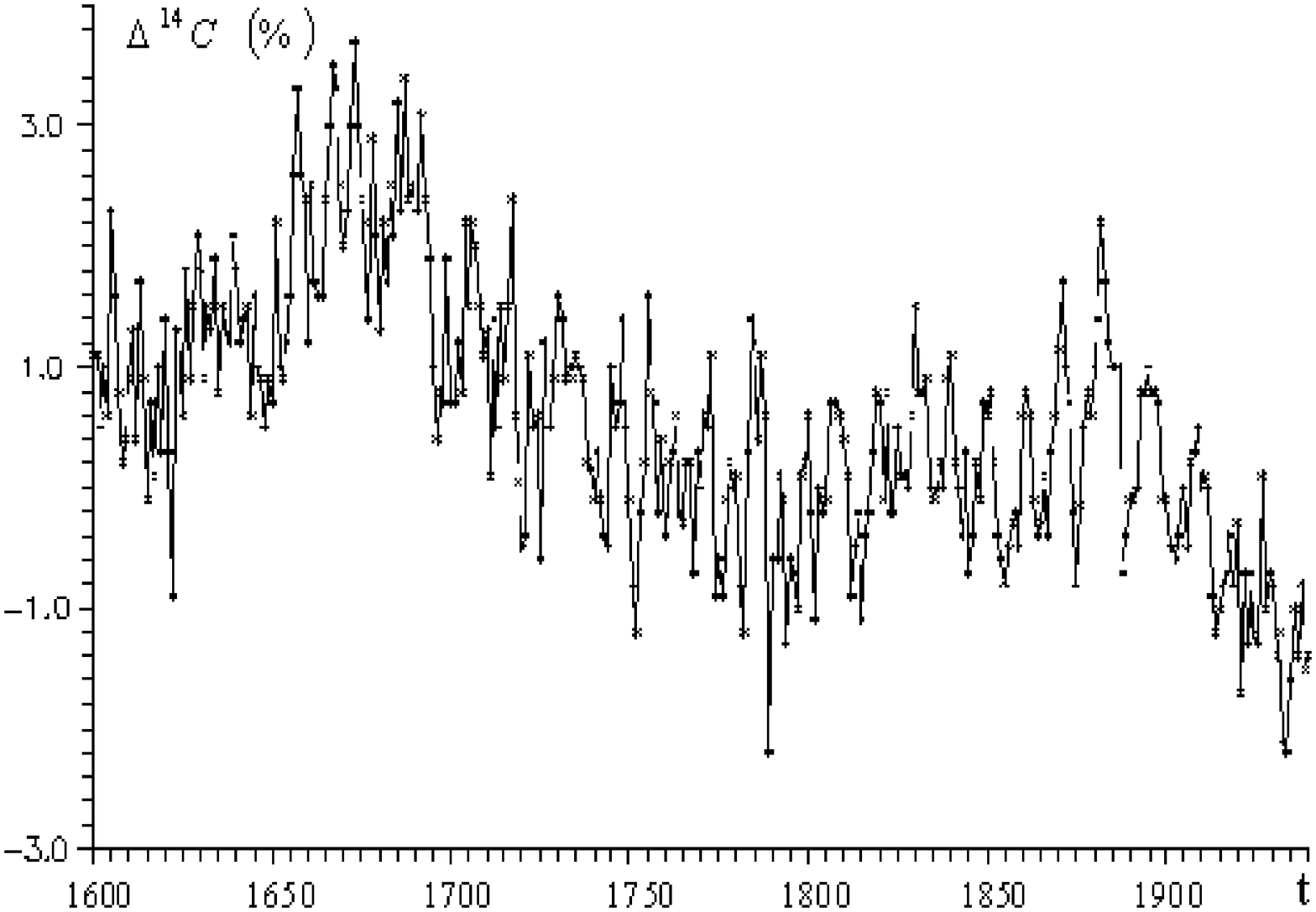}
\caption [1] {$^{14}C$ concentration in the atmosphere during
1600-1940 year.}
\end{center}
\end{figure}

\begin{figure}
\begin{center}
\includegraphics[angle=270, width=\textwidth  ]{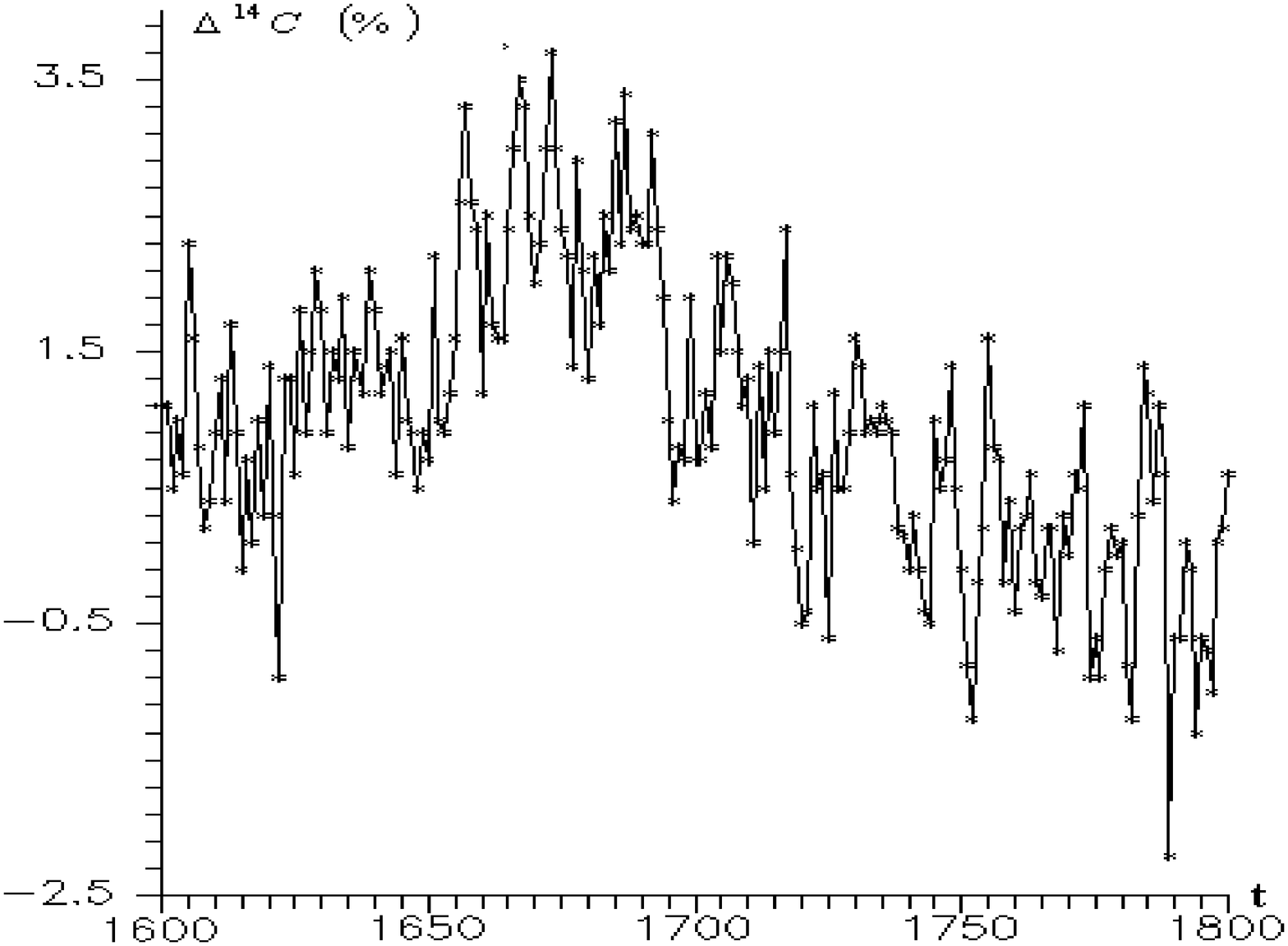}
\caption [1] {$^{14}C$ Concentration in the atmosphere during
1600-1800 year.}
\end{center}
\end{figure}

\begin{figure}
\begin{center}
\includegraphics[angle=0, width=\textwidth ]{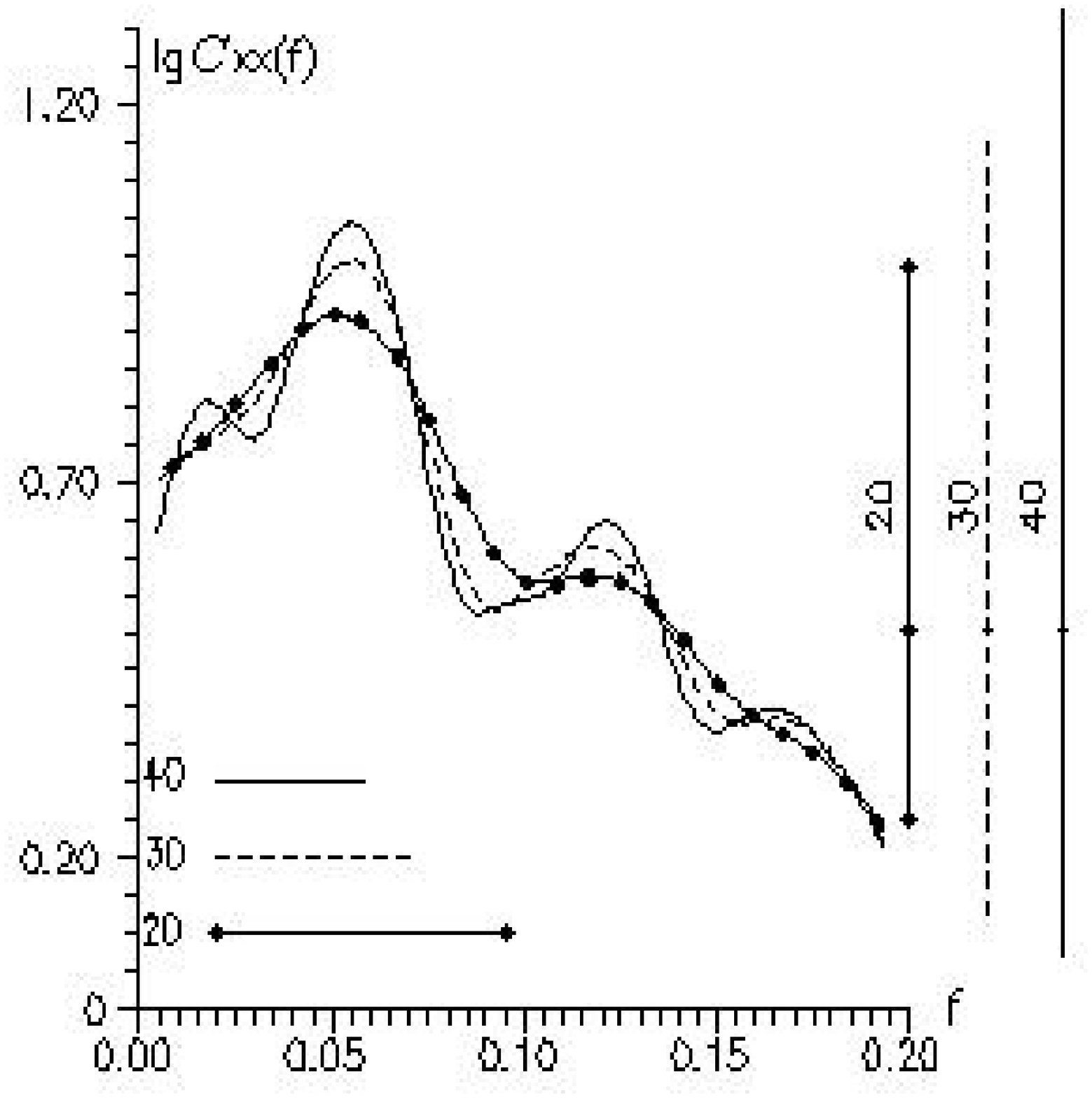}
\caption[ 2]{Dependence   of  spectral power $lg(C_{xx})$ on
frequency. For radiocarbon concentration in 1645-1715 yy.}
\end{center}
\end{figure}

\end{document}